%% file: main.tex
\documentclass[10pt,twocolumn,letterpaper]{article}

\usepackage{iccv}              % To produce the CAMERA-READY version
\input{preamble}

\definecolor{iccvblue}{rgb}{0.21,0.49,0.74}
\usepackage[pagebackref,breaklinks,colorlinks,allcolors=iccvblue]{hyperref}

\def\paperID{11733} % *** Enter the Paper ID here
\def\confName{ICCV}
\def\confYear{2025}

\title{CAGE-GS: High-fidelity Cage Based 3D Gaussian Splatting Deformation}

\author{Yifei Tong$^{1,}$\thanks{Equal contributions to this work.}\quad
\and
Runze Tian$^{1,}$\footnotemark[1]\quad
\and
Xiao Han$^{1,}$\footnotemark[1]\quad
\and
Dingyao Liu$^{1}$\quad
\and
Fenggen Yu$^{2}$\quad
\and
Yan Zhang$^{1}$\quad
\and
$^{1}$Nanjing University\quad 
$^{2}$Simon Fraser University\\
{\tt\small yifeitong754@outlook.com}
}

\begin{document}

\twocolumn[{%
\renewcommand
\twocolumn[1][]{#1}%
\maketitle
\centering
\includegraphics[width=0.9\linewidth]{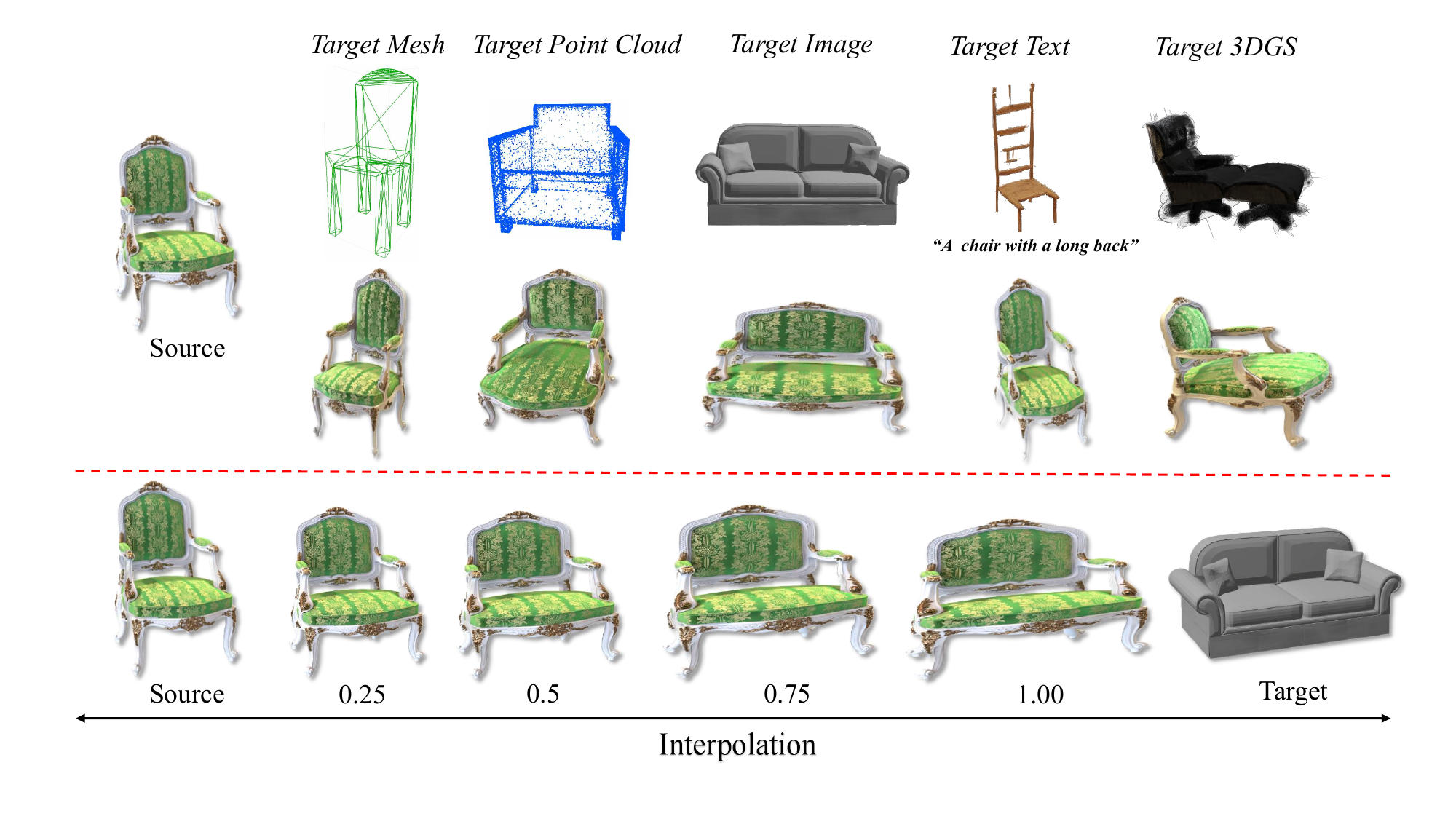}
\captionof{figure}{
    \textbf{\textit{Results of CAGE-GS.}} \textit{Top}: Our method deforms the source chair to match the target models, supporting multiple target representations, including meshes, point clouds, images, texts and 3DGS. \textit{Bottom}: Our method is able to control the deformation magnitude through cage interpolation.
\vspace{1em}
}
\label{fig:teaser}
}]
\begingroup
\renewcommand\thefootnote{}\footnote{*Equal contributions to this work.}%
\addtocounter{footnote}{-1}
\endgroup

\input{sec/0_abstract}    
\input{sec/1_introduction}
\input{sec/2_relatedwork}
\input{sec/3_method}

\input{sec/4_experiments}

\input{sec/5_conclusion}

{
    \small
    \bibliographystyle{ieeenat_fullname}
    \bibliography{main}
    
}

\end{document}

%% file: preamble.tex
\usepackage[dvipsnames]{xcolor}
\usepackage{comment}
\usepackage{capt-of}

\definecolor{green}{rgb}{0, 0.5, 0}
\definecolor{orange}{rgb}{0.8, 0.6, 0.2}
\definecolor{red}{rgb}{1.0, 0.0, 0.0}
\definecolor{teal}{rgb}{0.0, 0.4, 0.4}
\definecolor{purple}{rgb}{0.65,0,0.65}
\definecolor{saffron}{rgb}{0.95,0.75,0.2}
\definecolor{turquoise}{rgb}{0.0,0.5,0.5}
\definecolor{black}{rgb}{0.0, 0.0, 0.0}
\definecolor{gray}{rgb}{0.5, 0.5, 0.5}

\usepackage{float}
\usepackage{stfloats}

%% file: sec/0_abstract.tex
\begin{abstract}
\label{abstract}
As 3D Gaussian Splatting (3DGS) gains popularity as a 3D representation of real scenes, enabling user-friendly deformation to create novel scenes while preserving fine details from the original 3DGS has attracted significant research attention. We introduce CAGE-GS, a cage-based 3DGS deformation method that seamlessly aligns a source 3DGS scene with a user-defined target shape. Our approach learns a deformation cage from the target, which guides the geometric transformation of the source scene. While the cages effectively control structural alignment, preserving the textural appearance of 3DGS remains challenging due to the complexity of covariance parameters. To address this, we employ a Jacobian matrix-based strategy to update the covariance parameters of each Gaussian, ensuring texture fidelity post-deformation. Our method is highly flexible, accommodating various target shape representations, including texts, images, point clouds, meshes and 3DGS models. Extensive experiments and ablation studies on both public datasets and newly proposed scenes demonstrate that our method significantly outperforms existing techniques in both efficiency and deformation quality.

\end{abstract}

%% file: sec/1_introduction.tex
\section{Introduction}
\label{sec1:introduction}

Recent advancements in 3D Gaussian Splatting (3DGS)~\cite{kerbl20233d} have revolutionized real-time 3D rendering, offering an efficient representation for high-quality novel view synthesis and image-based 3D reconstruction. However, generating a new 3DGS model typically requires extensive multi-view images capturing, Structure-from-Motion (SfM)~\cite{snavely2006photo} processing and optimization, making it impractical for rapid content creation. As a result, researchers are shifting their focus toward editing or deforming existing 3DGS models to efficiently generate novel 3D scenes. 

Cage-based deformation~\cite{joshi2007harmonic, ju2005mean, lipman2008green, yifan2020neural, peng2022cagenerf, xie2024sketch,jiang2024vr} is a powerful technique in computer graphics that enables intuitive, structure-aware shape manipulation while preserving fine details. Unlike direct vertex-based deformation, which can lead to unstructured distortions, a cage provides a high-level control structure that smoothly influences the deformation of the enclosed model. Traditional cage-based deformation methods~\cite{joshi2007harmonic, ju2005mean, lipman2008green} often require manual manipulation of cage vertices, which can be time-consuming and unintuitive, especially for complex shapes. Recently, NeuralCage~\cite{yifan2020neural} overcomes this limitation by automatically learning the deformation mapping between a source and target model, eliminating the need for manual adjustments. However, these cage-based deformation methods primarily focus on mesh or implicit representations and are incompatible with the complexity and discreteness nature of 3DGS. 

High-fidelity deformation of 3DGS models remains a significant challenge, especially when aiming to transfer complex shape deformations while preserving fine texture details. Existing 3DGS deformation approaches either fail to maintain geometric accuracy or suffer from texture distortions due to improper handling of Gaussian parameters~\cite{wu2024gaussctrl, zhuang2024tip, cai2024dynasurfgs, huang2024sc, gao2024real}. 

To address these challenges, we introduce CAGE-GS, a cage-based 3DGS deformation method that seamlessly integrates cage-based shape control with Gaussian parameter optimization. Our approach leverages NeuralCage learning~\cite{yifan2020neural} to establish a structured deformation space, allowing automatic geometric transfer from a target model without requiring manual vertex manipulation. Specifically, our method binds the source cage with the centers of Gaussians from the source model. As the cage deforms, this binding drives the position change of the Gaussians. Additionally, we introduce a Jacobian matrix~\cite{spivak1965modern} based adjustment mechanism to dynamically update Gaussian covariance matrices, ensuring high-fidelity texture after deformation.

Our method enables intuitive cage-based deformation and leverages real-time 3DGS rendering to generate high-fidelity deformed 3DGS models efficiently. It also supports various model types, such as texts, images, point clouds, meshes and 3DGS models as target models for more practical applications. To summarize, our contributions include:
\begin{itemize} 

\item We introduce a cage-based 3DGS deformation method, allowing automated, structure-aware transformations without manual control point adjustments. 

\item We utilize the Jacobian matrix driven Gaussian parameter optimization, ensuring high-fidelity texture preservation after deformation.

\item Our 3DGS deformation method supports multiple input formats from users, including texts, images, point clouds, meshes and 3DGS models, making it broadly applicable for scene editing and content creation.

\item Extensive experiments on both public datasets and self-collected scenes demonstrate that our method outperforming existing techniques in both efficiency and deformation quality.
\end{itemize}

%% file: sec/2_relatedwork.tex
\section{Related Work}
\label{sec2:relatedwork}

\noindent
\textbf{Cage-based Deformation.} Cage-based deformation (CBD) is a widely used technique for deforming 3D meshes by transforming the space enclosed within a coarse control cage and applying the computed transformation to the mesh vertices. The effectiveness of CBD hinges on the definition of cage coordinates, which determine how interior points interpolate deformations from cage vertices. Various coordinate formulations~\cite{ju2005mean, joshi2007harmonic, lipman2008green, xian2012automatic, sacht2015nested, calderon2017bounding,thiery2012cager} have been proposed, leading to advancements in mesh deformation. CBD is particularly valued for its ability to provide intuitive, low-dimensional control over complex deformations while preserving mesh details and preventing self-intersections. 
Recent works extend CBD principles to neural implicit representations. DeformingNeRF~\cite{xu2022deforming} and NeRFShop~\cite{jambon2023nerfshop} extend CBD to NeRF~\cite{mildenhall2021nerf}, where model deformation is achieved by modifying the sampled points along camera rays during volume rendering. However, adapting these methods to 3DGS is challenging, which relies on rasterization rather than volume rendering.
To adapt CBD to 3DGS, GSDeformer~\cite{huang2024gsdeformerdirectrealtimeextensible} constructs cages using boundary proxies, directly converting the density field represented by 3DGS into a voxel grid for deformation. Similarly, D3GA~\cite{zielonka2023drivable} places Gaussians inside a cage and adjust their mean positions based on cage vertex movements, achieving deformation of avatars. Despite their success, these methods lack mechanisms for deformation transfer, making it difficult to apply learned deformations from one instance to another.

\noindent
\textbf{3D Gaussian Splatting Editing.} With the growing popularity of 3DGS in computer vision, some research has emerged to support 3DGS editing. Text-guided editing methods~\cite{wang2024gaussianeditor, wu2024gaussctrl, zhuang2024tip, chen2024gaussianeditor, palandra2024gsedit, wang2024view} allow users to modify 3DGS models with simple text prompts. However, these methods often lead to semantic changes, limiting their ability to precisely edit geometry and texture details.
Another category of editing methods~\cite{huang2024sc, cai2024dynasurfgs, ma2024reconstructing} enables dynamic geometric deformations by leveraging sufficient prior knowledge extracted from monocular videos. However, they require additional video supervision and are highly sensitive to the camera viewpoints provided in the dataset, limiting their applicability. Additionally, drag-based editing methods~\cite{chen2024mvdrag3d, shen2024draggaussian} project user’s drag operations onto 2D images and optimize the 3D scene based on the edited images. Despite achieving consistent results, they require significant parameter fine-tuning, and the dragging points can sometimes become occluded in all views.
In contrast, our method directly deforms 3DGS models by cage-based deformation transfer with a target shape given from the user, a capability lacking in existing methods. This significantly broadens the potential applications of 3DGS.

\noindent
\textbf{Shape Deformation Transfer.} 
In digital content creation, achieving specific deformation effects is often crucial for applications such as dataset augmentation and animation synthesis. Shape deformation transfer~\cite{sumner2004deformation, baran2009semantic, ben2009spatial, hanocka2018alignet, wang20193dn, gao2019sdm, sung2020deformsyncnet, yifan2020neural, peng2022cagenerf, groueix2019unsupervised, huang2017learning} aims to align the shape of a source model with a target model while preserving structural integrity and texture details. A key challenge in the process of deformation inference is establishing a correspondence between the shapes of the source and target models. This can be achieved either by explicitly inferring corresponding points~\cite{ben2009spatial, sumner2004deformation}, or by implicitly adjusting the deformation field based on the latent encoding of the target shape~\cite{wang20193dn, hanocka2018alignet}.
However, these methods struggle to preserve surface details and only perform well when the input shape closely matches the target shape.

NeuralCage~\cite{yifan2020neural} and CageNeRF~\cite{peng2022cagenerf} introduce frameworks that parameterize deformation spaces using cage structures, enabling shape transfer through implicit deformation field adjustments. However, these methods rely on mesh and NeRF-based 3D representations, making them incompatible with 3DGS, where deformation requires modifying parameters such as the covariance matrices of Gaussians. To overcome this challenge, we integrate the cage-based deformation framework into 3DGS, allowing it to optimize the parameters of Gaussians, achieving deformation transfer for 3DGS models. Our method provides a convenient and high-fidelity solution for deformation transfer, leveraging the efficient model creation ability of 3DGS.

%% file: sec/3_method.tex
\section{Method}
\label{sec3:method}
\begin{figure*}[t]
  \centering
  \includegraphics[width=1\linewidth]{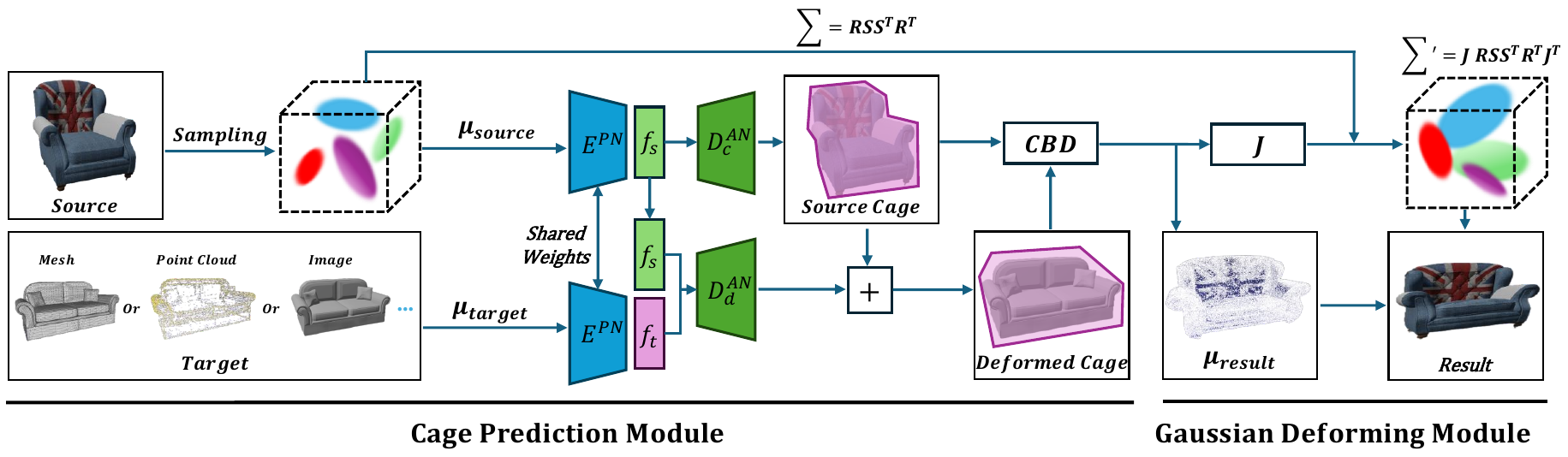}
  
   \caption{\textbf{\textit{Overview.}} Our cage-based deformation framework has two modules. In the Cage Prediction Module, it learns the source cage from the source model and the cage offset from the target to generate a deformed cage. In the Gaussian Deforming Module, the deformed cage and optimized mean value coordinates from the source cage are used to deform the positions of the 3D Gaussians while keeping their covariance matrix intact. Finally, we employ the Jacobian matrix to effectively update the covariance matrix of the 3D Gaussians, maintaining high-fidelity texture on the rendered images. Our method supports multiple target representations, including texts, meshes, point clouds, images and 3DGS.}
   \label{fig:overview}
\end{figure*}

Given a set of multi-view images, we express its geometry and appearance as a 3DGS model. Additionally, we provide a target shape, whose representation can take the form of texts, images, meshes, point clouds and 3DGS. Our goal is to deform the source 3DGS model to match the geometry of the target model while ensuring its surface texture details are fully preserved. An overview of our method is shown in ~\cref{fig:overview}.
In the following, we first introduce some preliminaries, including the 3DGS representation and cage-based deformation in ~\cref{sec3.1:preliminary}. Then, we present the cage prediction module in ~\cref{sec3.2:neural}, followed by Jacobi matrix based 3D Gaussians deformation in ~\cref{sec3.3:deform}.

\subsection{Preliminary}
\label{sec3.1:preliminary}
\textbf{3D Gaussian Splatting.} 3D Gaussian Splatting~\cite{kerbl20233d} is an advanced 3D representation that uses a set of anisotropic 3D Gaussians to represent the scene, denoted as $G=\{g_1, g_2, \dots, g_N\}$, where $g_i=\{\mu, \Sigma, c, \alpha\}$, $i\in\{1, \dots, N\}$. In this formulation, $\mu$ represents the center position of the Gaussian, $\Sigma$ is the 3D covariance matrix, $c$ is the color encoded with spherical harmonic coefficients, and $\alpha$ denotes opacity. To ensure that $\Sigma$ is positive semi-definite and allows independent optimization, it is decomposed as $\Sigma = RSS^{T}R^{T}$, where ${R}$ is a rotation matrix that can be easily converted to a quaternion $q$, and ${S}$ is a diagonal scaling matrix.

\noindent
\textbf{Cage-based Deformation.}
Cage-based deformation(CBD) is a type of freeform mesh deformations. Given a cage $C_s$ that encloses a mesh, the position of points $p \in R^3$ inside $C_s$ is represented as a weighted sum of cage vertices $v^s$:
\begin{equation}
\label{eq:source}
p = \sum_{i} \omega_i(p) v^s_i
\end{equation}
Here we choose mean value coordinates (MVC)~\cite{ju2005mean} as the weight function $\omega$ for its simplicity and differentiability. 
After $C_s$ is deformed to the new cage $C_{s\to t}$ with vertices $v^{s\to t}$, the deformed positon of $p$ is calculated as:
\begin{equation}
p' = \sum_{i} \omega_i(p) v^{s\to t}_i
\label{eq:deform}
\end{equation}

\subsection{Learning Cages for Deformation}
\label{sec3.2:neural}

Manual manipulation of the source cage to generate the deformed cage can be time-consuming and unintuitive, especially when the cage has redundant vertices for complex shapes. We draw inspiration from NeuralCage~\cite{yifan2020neural}, combining traditional cage deformation techniques with advanced neural network architectures to learn the deformed cage from a given target shape. We represent spatial points on the source model as a weighted average of the cage vertices using generalized barycentric coordinates for deformation. As shown in Figure~\ref{fig:overview}, the neural network consists of two encoders $E^{PN}$ with shared weights, one decoder $D^{AN}_c$ and one decoder $D^{AN}_d$. The input of the network is two point clouds sampled from the source and target models separately, while the output is a deformed source cage.

In testing phase, We first randomly sample $N_{sample}$ (set to 30,000) Gaussians from the source 3DGS model, and use their central positions $P_{sample}$, as sampled points input into the source encoder. For the input target model, we then adopt different strategies depending on its representation form. For 3DGS representation, we follow the same method as described above to sample the input point cloud. For text and image input, we first use a learning-based method to generate a rough 3D proxy model as the target. Then, we sample $N_{sample}$ points from the proxy surface as the target input point cloud. For mesh and point cloud data, we directly sample point cloud on the model as input. 
 
The encoders of the network output $f_s$ and $f_t$, representing the encoded feature of the input and target point cloud. The decoder $D^{AN}_c$, after receiving $f_s$, decodes the source model's cage $C_s$. The decoder $D^{AN}_d$, based on $f_s$ and $f_t$, predicts the cage offset for $C_s$, yielding a deformed cage $C_{s\to t}$. The function of the network can be expressed by the below equation:
\begin{align}
    f_s &=E^{PN}(S_s) ,f_t =E^{PN}(S_t)\\
    C_s &=D^{AN}_c(f_s)+C_0 \\
    C_{s\to t} &=D^{AN}_d(f_t,f_s)+C_s 
\end{align}
where $S_s$ and $S_t$ are the source and target point clouds input into the network. Note that the source cage is represented by the template cage $C_0$ and its predicted offset $D^{AN}_c(f_s)$.

To make the deformed cage well-aligned with the target model and generalized to novel shapes, we remove the pre-training process in NeuralCage and perform the optimization process per model. The cage prediction module is optimized with the source and target point clouds during testing, and the loss is the same as NeuralCage to encourage positive mean value coordinates, alignment to target and surface normal preservation.

%The deformed cage $C_{s\to t}$ is then used to guide the deformation of the 3D Gaussians.}

\subsection{Cage-based 3D Gaussians Deformation}
\label{sec3.3:deform}
 
After optimizing the neural network in ~\cref{sec3.2:neural}, we can obtain the optimized mean value coordinates and the deformed cage $C_{s\to t}$. Then we can use ~\cref{eq:deform} to update the central position $\mu$ of 3DGS and obtain $\mu'$ for the deformed Gaussian. However, the covariance matrix $\Sigma$ of the 3DGS has not been changed yet. Since the covariance matrix encodes anisotropic Gaussian splats, an incorrect update would lead to blurring, stretching, or distortion of textures after deformation, as shown in ~\cref{fig:Jacobi_matrix}. To counter this, the Jacobian matrix~\cite{spivak1965modern} is used to approximate the local deformation and correctly transform the Gaussian covariance matrices. 

% To address this issue, we establish a normalized deformation space by Jacobian matrix to obtain the new covariance matrix $\Sigma$ of the deformed Gaussian because \fg{[FG: why do we use Jacobian matrix initially, what is its advantage? update here]}Jacobian matrix represents local linear approximation of the deformation. This allows us to efficiently adjust parameters such as the rotation and scaling matrix for each Gaussian.

%To adjust the covariance matrix $\Sigma$, we can use a local linear approximation of the deformation. 
By evaluating the Jacobian matrix at the center $\mu$, we can estimate the covariance matrix for each Gaussian. However, calculating the Jacobian matrix for each Gaussian is computationally expensive. Therefore, we sample $m$ Gaussians from the total $N$ Gaussians and compute the Jacobian matrix for the sampled Gaussians first. The specific calculation is as follows:

\begin{align}
%\mu'_{\text{sample}} &= \text{CBD}(\mu_{\text{sample}}) \notag \\
J &= \frac{\partial \mu'_{\text{sample}}}{\partial \mu_{\text{sample}}}
\end{align}

where $\mu_{\text{sample}}$ represents the sampled Gaussian centers for Jacobian matrix calculation, and $\mu'_{\text{sample}}$ denotes their deformed positions.

After computing the Jacobian matrix, we can obtain the deformed covariance matrix $\Sigma'$ as follows:
\begin{equation}
\Sigma' = JRSS^{T}R^{T}J^T
\end{equation}

By decomposing the deformed covariance matrix, we can extract the new rotation matrix $R'$ and scaling matrix $S'$ of each deformed Gaussian, as shown below:
\begin{align}
    R'S'S'^{T}(R')^{T} &= \text{SVD}(\Sigma'), \\
    S' &= \sqrt{S'S'^T}
\end{align}

For the unsampled Gaussians, we use k-nearest neighbors (kNN) to find the nearest sampled Gaussian and assign its Jacobian matrix to the unsampled Gaussians. Once the Jacobian matrix is assigned, the covariance matrix update of these unsampled Gaussians can be achieved same as above. Through ~\cref{sec3.2:neural} and ~\cref{sec3.3:deform}, we can successfully obtain the key parameters of the deformed Gaussian: the Gaussian center $\mu'$, new rotation matrix $R'$, and new scaling matrix $S'$. This method accurately transforms the covariance matrix to maintain anisotropic properties of Gaussians and ensures that the relative arrangement of Gaussians remains coherent after deformation, preventing unwanted artifacts.

%% file: sec/4_experiments.tex
\section{Experiments}
\label{sec4:experiments}

\subsection{Experimental Setup}
\label{sec4.1:setup}

\textbf{Dataset.}
The test objects are collected from ShapeNet~\cite{chang2015shapenet}, generated by Gaussian Anything~\cite{lan2024gaussiananything} and real-captured by ourselves. For the objects in ShapeNet, we use the render scripts in ~\cite{ma2024shapesplat} to transform them into 3DGS datasets.

\noindent
\textbf{Implementation Details.}
We conduct our experiments on a single NVIDIA RTX 3090 GPU. The size of the source model ranges from 100k to 500k Gaussians. Through the entire process, we use 64-bit floating-point precision to minimize numerical errors. To training the cage network, we sample 30,000 points and train for 50 epochs. During the CBD process, we use batch processing to reduce the memory requirement of CBD. The Gaussian centers are grouped with a batch size of 30,000. When computing the Jacobian matrix, the default number of sampled points is set to 10,000. To accelerate the Jacobian matrix computation using PyTorch's automatic differentiation, we process the sampled points in batches of 200.

% \noindent
% \textbf{Baselines.}

\noindent
\textbf{Running Time.}
Our deformation time cost depends on the number of source Gaussians. For a source model containing 200k Gaussians, our deformation process takes approximately 8 minutes, which includes the time for cage network optimization, Jacobian matrix computation, CBD solving, and the application of the Jacobian matrix for covariance transformation.

\noindent
\textbf{Evaluation Metrics.}
The metrics we use to evaluate the performance are Chamfer Distance(CD) and DINO directional cosine similarity (DINO). We calculate CD between the deformed model and the target model. And DINO is defined as:
$DINO=\cos(f_{tgt}-f_{src},f_{dfm}-f_{src})$, where $f_{tgt}$, $f_{src}$ and $f_{dfm}$ are features extracted from target, source and deformed models respectively, using the DINOv2\cite{oquab2023dinov2} model on each view.
Given that deformation quality is highly dependent on user perception, we conduct a user study to evaluate subjective preferences, collecting votes from 60 participants. The deformation results from different methods are randomly placed in each question.

% We use Chamfer Distance(CD) to measure the geometry distance between the deformed model and the target model.

\begin{figure}[!h]
  \centering
  \includegraphics[width=1\linewidth]{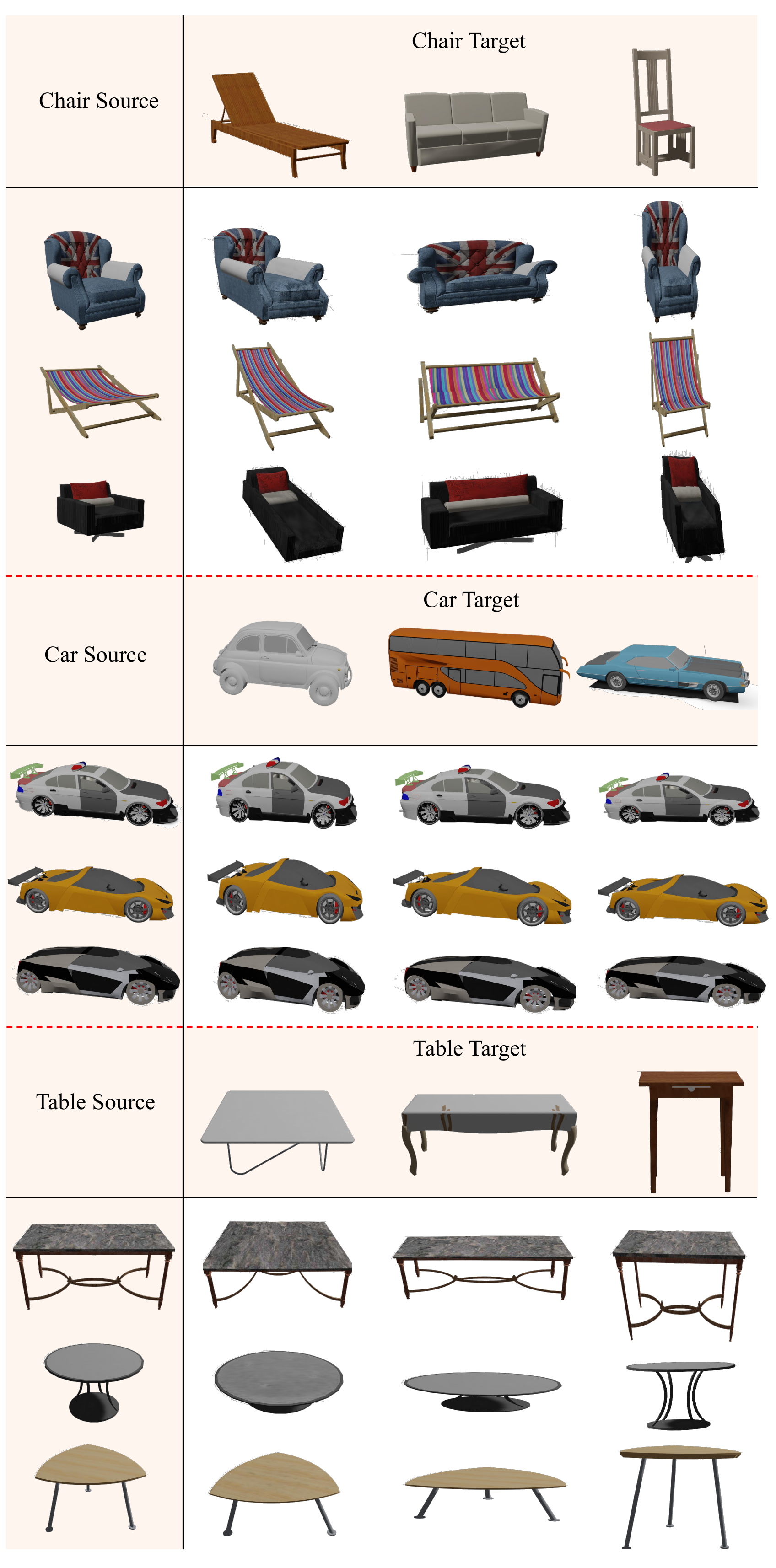}
  
   \caption{\textbf{\textit{Deformation results on ShapeNet dataset.}} We validate our method on the chair, car and table categories from ShapeNet~\cite{chang2015shapenet}. The results show that our method aligns the source shape with the target shape while preserving the source texture.}  
   \label{fig:shapenet_res}
\end{figure}

\subsection{Visual Results} 
\textbf{Category-specific 3DGS deformation.} 
As shown in ~\cref{fig:shapenet_res}, our method can learn deformation patterns among multiple shapes within the same category from the dataset, enabling model creation with the desired target shape. Even though our training process omits all semantic supervision, such as part annotations, the transformations remain reasonable without feature distortion. Moreover, both texture and geometry details are well preserved.

\begin{figure*}[thbp]
  \centering
  \includegraphics[width=1\linewidth]{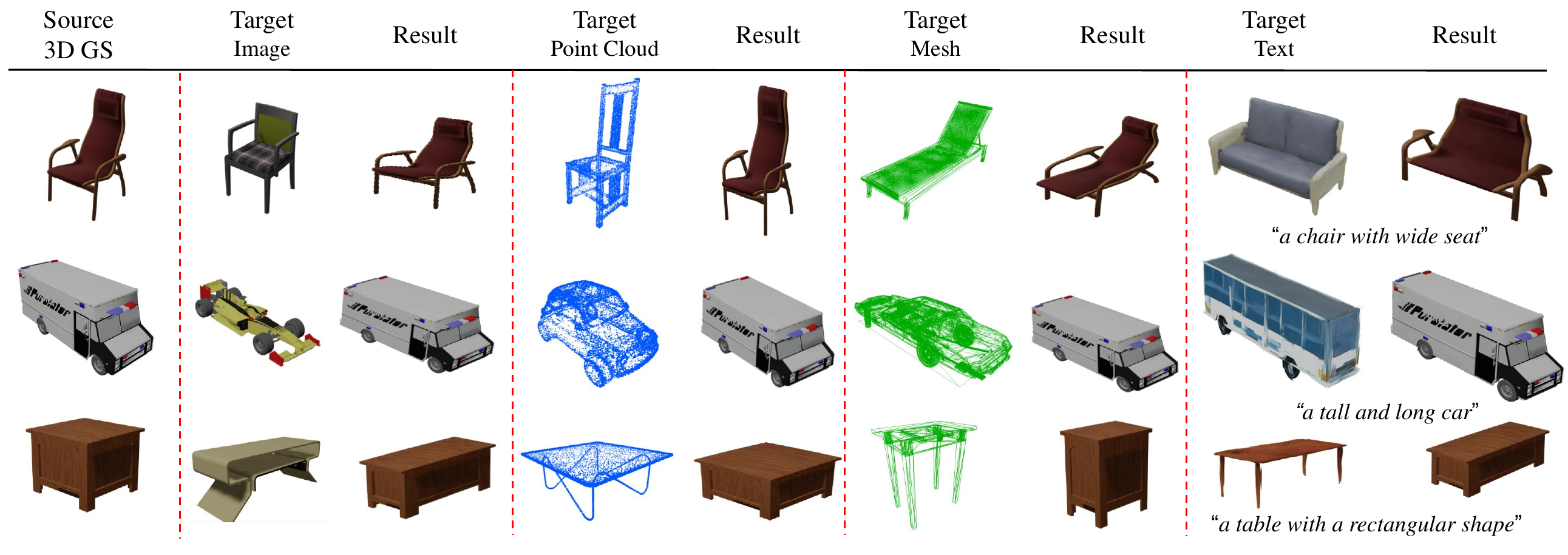}
  
   \caption{\textbf{\textit{Deformation results on different target representations.}} The results show that our method can support different target formats, including images, point clouds, meshes and 3DGS model generated with text prompt.}
   
   \label{fig:target}
\end{figure*}

\noindent
\textbf{Using different representations as targets.}
To validate the compatibility of our method, we use different  representation forms as targets, including single images, point clouds, meshes and 3DGS model generated with text prompt from GaussianAnything~\cite{lan2024gaussiananything}. The quality of the generated 3DGS from GaussianAnything is quite low, but it can provide users with a rough model, which we can then use as the target for deformation. If the target is a single image, we use AtlasNet ~\cite{groueix2018} to reconstruct a target mesh. Despite the poor quality of the proxy, it still serves as a valid target for our method. For mesh and point cloud target, we directly sample on the target to construct the input of our neural network. As shown in ~\cref{fig:target}, regardless of the representation form used for the target model, even when the target proxy possesses coarse geometric features and contains multiple defects, our method can still achieve proper deformation results.

\begin{figure}[!h]
  \centering
  \includegraphics[width=1\linewidth]{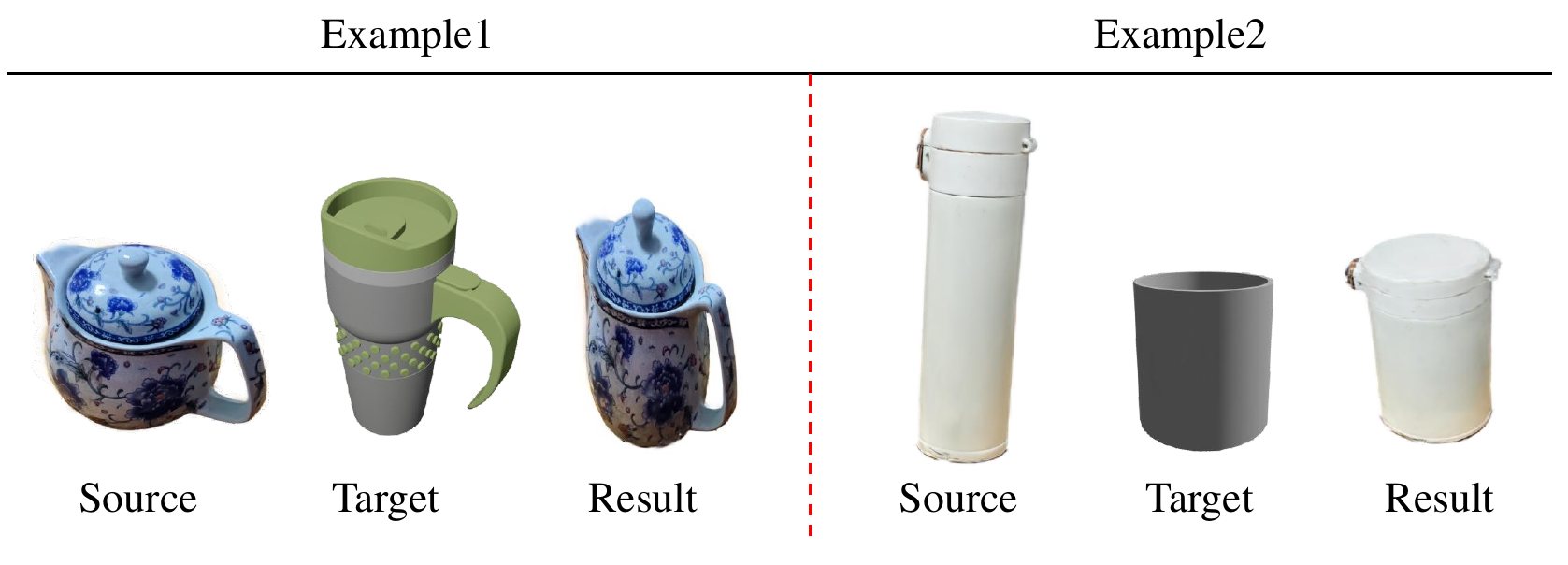}
  
   \caption{\textbf{\textit{Deformation results on real-captured data.}} We validate our method on a real-captured teapot and a mug. The results show that our method can apply deformation transfer to the real-captured models. }
   
   \label{fig:real_data}
\end{figure}

\noindent
\textbf{Using real-captured data as source model.}
To validate the applicability of our proposed method, we conduct experiments on real-captured data. We use a cellphone to take multi-view images of the object and use 3DGS to build the source model. Then we apply our method to perform deformation transfer on the real-captured model. As shown in ~\cref{fig:real_data}, our method can easily apply the desired deformations to the source 3DGS models.

\begin{figure}[!h]
  \centering
  \includegraphics[width=1\linewidth]{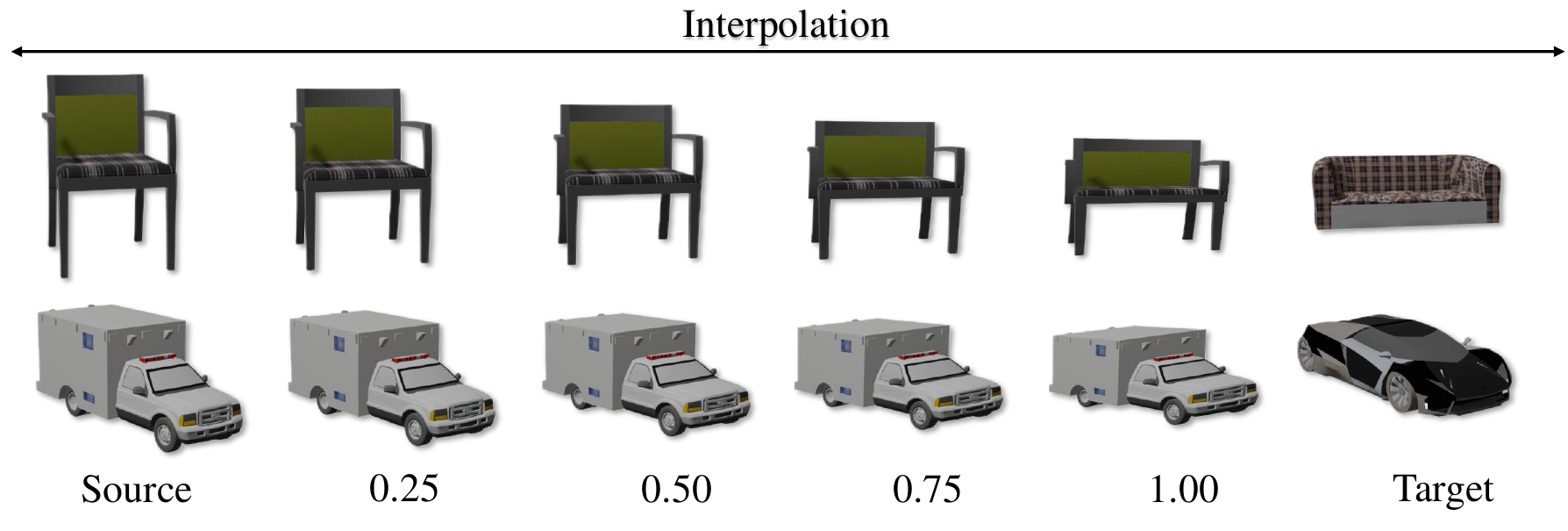}
  
   \caption{\textbf{\textit{Deformation magnitude control with interpolation.}} With cage interpolation, our method is able to control the deformation magnitude of the chair and the car.}
   \label{fig:cage_interpolation}
\end{figure}

\noindent
\textbf{Controlling deformation magnitude with cage interpolation.}
As shown in ~\cref{fig:cage_interpolation} , our method allows control over the deformation magnitude through linear interpolation of the cage. Given the source cage $C_s$, the deformed cage $C_{s\to t}$ and interpolation $\lambda$, we compute a new cage as $C_{new} = \lambda C_{s\to t} + (1-\lambda) C_s$.
Replacing $C_{s\to t}$ with $C_{new}$ in the CBD process produces a deformation result with the specified magnitude. A $\lambda$ closer to 1.0 results in a shape closer to the target model, while a $\lambda$ closer to 0.0 preserves more of the source model. In our experiments, we tested values such as $\lambda=0.25$, $\lambda=0.5$ and $\lambda=0.75$.

\begin{figure*}[thbp]
  \centering
  \includegraphics[width=1\linewidth]{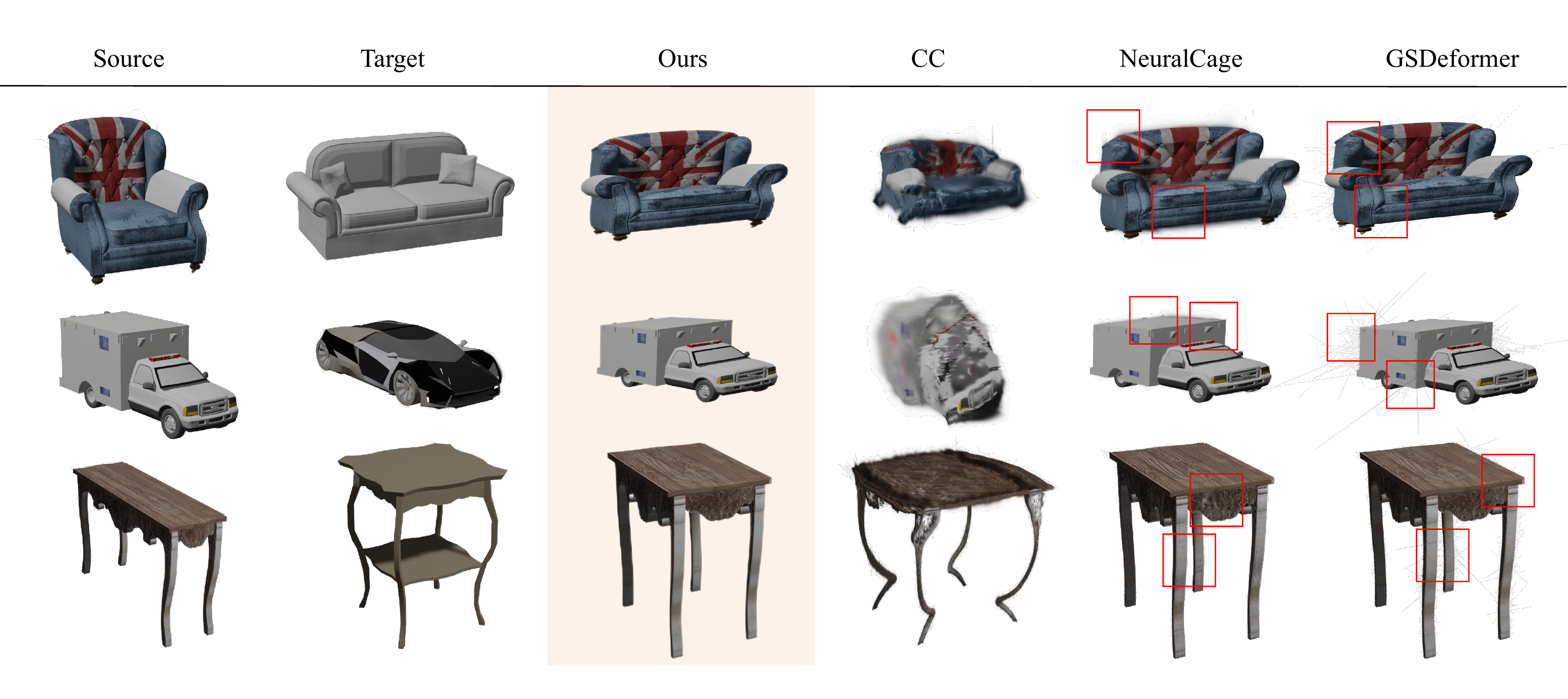}
  
   \caption{\textbf{\textit{Qualitative deformation comparison results.}} We compare our method with CC~\cite{groueix2019unsupervised}, NeuralCage~\cite{yifan2020neural} and GSDeformer~\cite{huang2024gsdeformerdirectrealtimeextensible} on the chair, car and table examples from ShapeNet ~\cite{chang2015shapenet}. The red boxes indicate areas where the deformation effect is poor, please zoom in to see the blurred area. The results show that our method can better preserve texture details after deformation.}
   
   \label{fig:comparison}
\end{figure*}

\noindent
\textbf{Qualitative comparison with baselines.}
We conduct qualitative comparisons with other deformation transfer methods~\cite{groueix2019unsupervised, yifan2020neural, huang2024gsdeformerdirectrealtimeextensible} in ~\cref{fig:comparison}.
Cycle-consistency(CC)~\cite{groueix2019unsupervised} introduces a cycle consistency loss when training a shape deformation network, which is then used to deform the source point cloud, bearing similarities to our task. We estimate the deformation of 3DGS by analyzing the changes of the source point clouds before and after deformation. As shown in ~\cref{fig:comparison}, although CC's deformation results share some resemblance with the target, it generates significant curvature and artifacts, leading to a decline in quality.
During the experiments, NeuralCage~\cite{yifan2020neural} and GSDeformer~\cite{huang2024gsdeformerdirectrealtimeextensible} also adopt CBD to update 3DGS centers, enabling geometric deformation. But without Jacobian matrix based covariance matrix updating, we observe they exhibit certain issues in preserving texture details after deformation. For instance, NeuralCage may produce blurring artifacts in texture rendering, and GSDeformer may result in elongated Gaussians, as shown in the red-boxed regions in ~\cref{fig:comparison}. For high-resolution and clearer images, please refer to the supplementary materials.

\subsection{Quantitative Results}
We use the evaluation metrics illustrated in ~\cref{sec4.1:setup}.
For user study, we anonymize our method and the baseline methods, providing users with the source model, the target model, and the deformation results rendered from three different viewpoints. Users are asked to select the result that best aligns with their preference. 
The quantitative comparison results are shown in ~\cref{tab:metrics}. Our method has similar CD value with NeuralCage~\cite{yifan2020neural} and GSDeformer~\cite{huang2024gsdeformerdirectrealtimeextensible} because we all use CBD to update 3DGS centers. Despite this, our method achieves significantly better scores on DINO and user preference. The results demonstrate that our method can better preserve surface texture while enabling geometry deformation.

\begin{table*}[!htb]
    \centering
    \begin{tabular*}{\textwidth}{@{\extracolsep{\fill}}c|ccccc}
    \toprule
    \ &CC\ &NeuralCage\ &GSDeformer\ &Ours\ \\
    \midrule
    CD($\downarrow$)\ &0.1622\ &0.0998\ &0.0998\ &\textbf{0.0997}\ \\
    DINO($\uparrow$)\ &0.367\ &0.385\ &0.374\ &\textbf{0.402}\ \\
    User Votes($\uparrow$)\ &3.3\%	\ &11.7\% \ &21.7\% \ &\textbf{63.3\%}\ \\
   
    \bottomrule
    \end{tabular*}
    \caption{\textbf{\textit{Quantitative evaluation.}} We compare our method with CC~\cite{groueix2019unsupervised}, NeuralCage~\cite{yifan2020neural} and GSDeformer~\cite{huang2024gsdeformerdirectrealtimeextensible}. We present the average CD/DINO values and the user votes on the deformation results from ShapeNet~\cite{chang2015shapenet}.}
    \label{tab:metrics}
\end{table*}

\subsection{Ablation Study}

\begin{figure}[!h]
  \centering
  \includegraphics[width=1\linewidth]{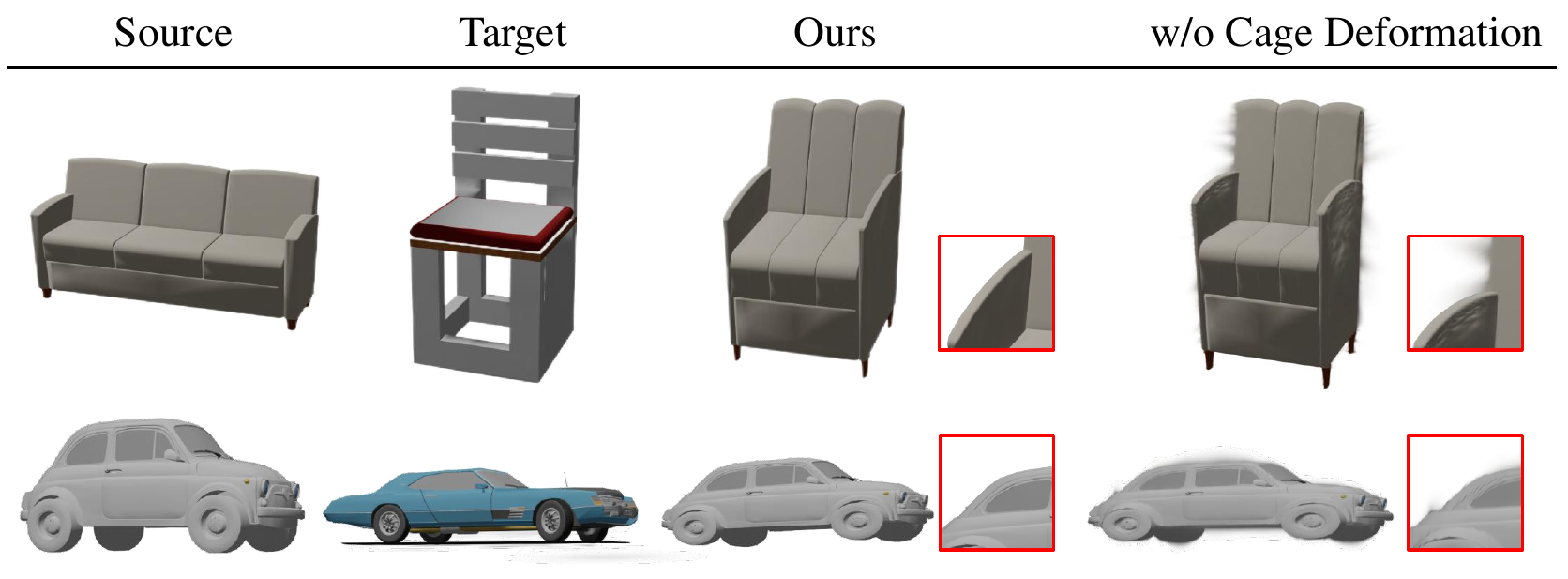}
  
   \caption{\textbf{\textit{Effectiveness of cage-based deformation.}} Using cage-based deformation and jacobian matrix to optimize Gaussian parameters can preserve texure details and reduce artifacts compared to anisotropic scaling.}
   \label{fig:cage_ablation}
\end{figure}

\textbf{Effectiveness of cage deformation.}
The most straightforward method of object deformation is anisotropic scaling, which adjusts the source model's bounding box to match the target model's bounding box. Specifically, we update the centers and the scale parameters of the source 3DGS model during anisotropic scaling. In ~\cref{fig:cage_ablation}, we show the comparison results of anisotropic scaling to our cage-based deformation technique. While anisotropic scaling preserves local structures, it fails in preserving high-fidelity texture, further validating the effectiveness of our cage-based deformation framework.

%While anisotropic scaling preserves local structures, it fails to account for the specific proportional changes required in different regions, highlighting the necessity of employing cage technology for optimal deformation in such cases. In contrast, our cage-based method demonstrates superior performance in matching corresponding semantic parts, further validating the effectiveness of our cage deformation framework.

\begin{figure}[!h]
  \centering
  \includegraphics[width=1\linewidth]{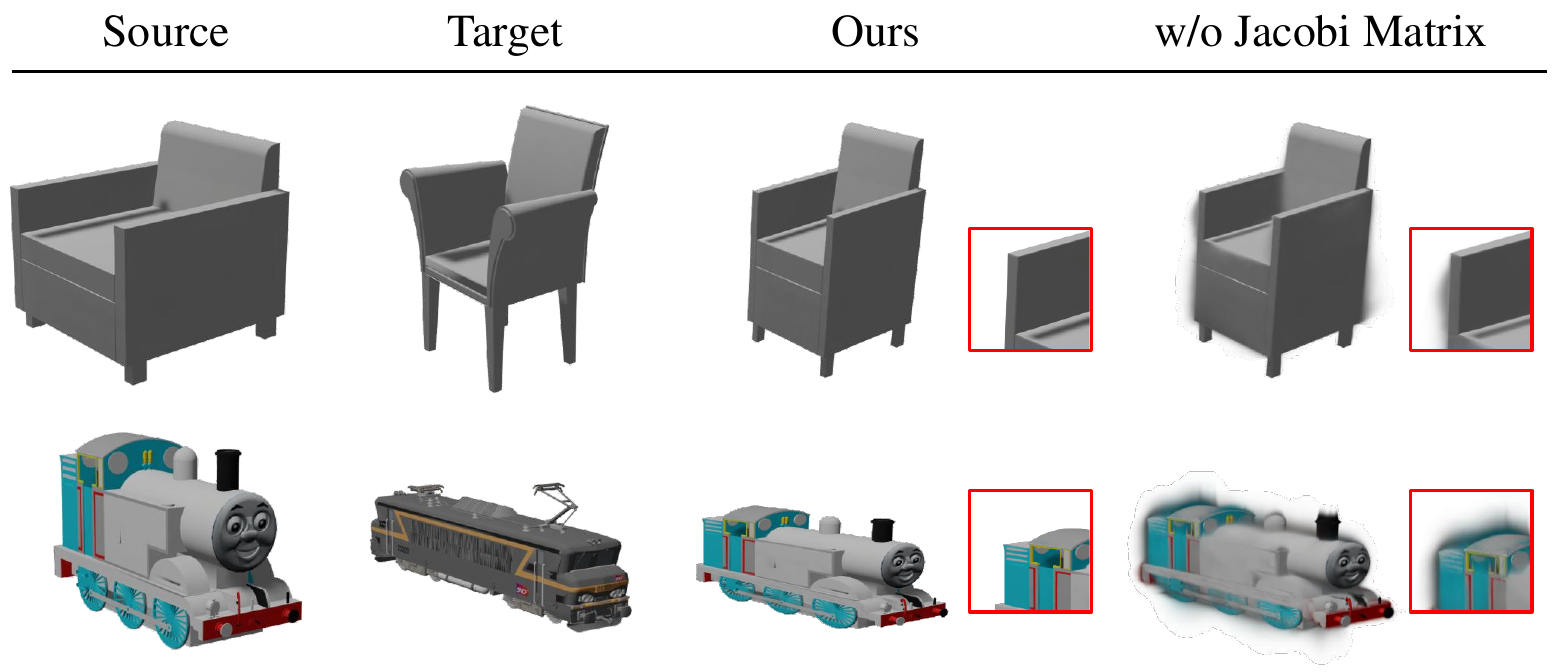}
  
   \caption{\textbf{\textit{Effectiveness of Jacobi matrix.}} Using Jacobi matrix to optimize Gaussian parameters can preserve texture details and reduce artifacts.}
   \label{fig:Jacobi_matrix}
\end{figure}

\noindent
\textbf{Effectiveness of Jacobi matrix.}
To validate the effectiveness of our method in updating Gaussian covariance parameters, we show results by adjusting only the position of Gaussians during cage-based deformation. As shown in ~\cref{fig:Jacobi_matrix}, our method not only maintains the geometric structure after deformation but also preserves the texture details more effectively. In contrast, adjusting only the position of Gaussians results in blurred textures and artifacts in the deformed model.

\begin{figure}[!h]
  \centering
  \includegraphics[width=1\linewidth]{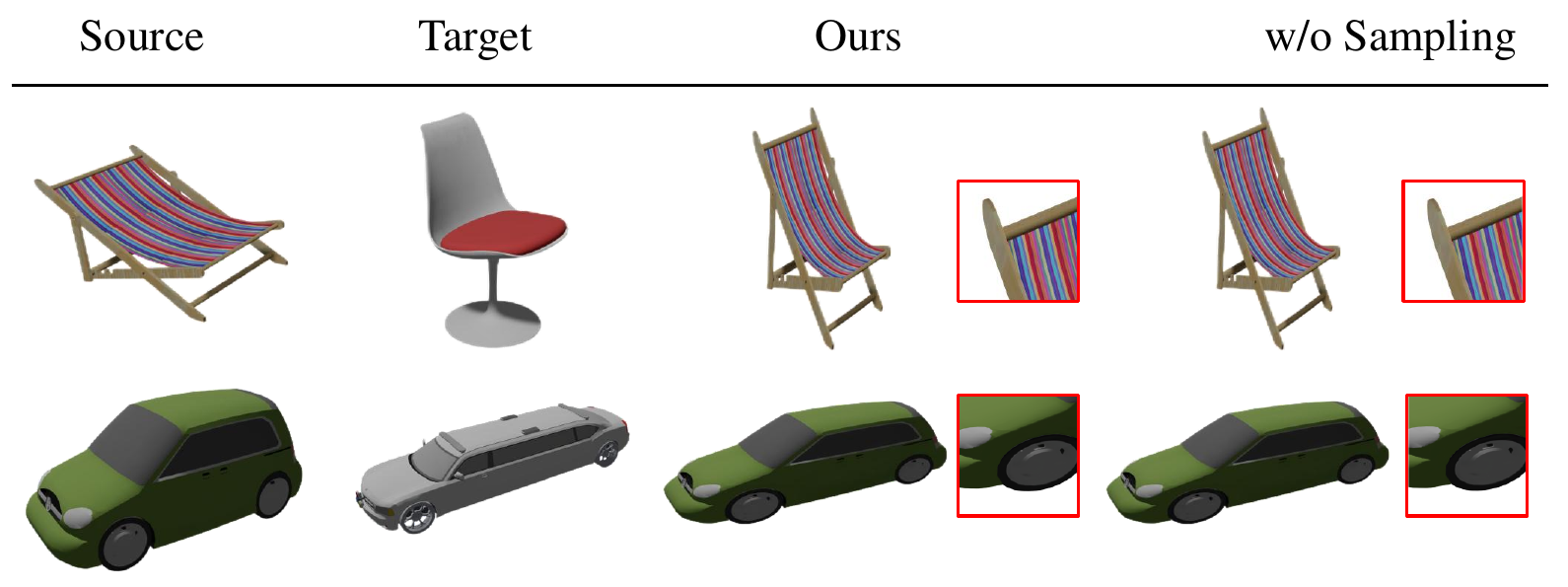}
  
   \caption{\textbf{\textit{Effectiveness of Sampling.}} Our sampling method can maintain the rendering quality while reducing time cost.}
   \label{fig:sampling}
\end{figure}

\noindent
\textbf{Effectiveness of sampling.}
We calculate the Jacobian matrix for down-sampled 3DGS in ~\cref{sec3.3:deform} and use kNN to update the remained 3DGS, aiming to reduce the time cost without compromising the rendering quality. Given the decisive impact of Jacobian matrix's accuracy on the final rendering result, we show the results without down-sampling for the Jacobian matrix computation, where we calculate the Jacobian matrix for each 3DGS. As shown in ~\cref{fig:sampling}, the sampling method barely affect rendering quality. And ~\cref{tab:time} shows that the deformation time cost is significantly reduced, greatly improving our method's efficiency.

\begin{table}[!h] 
    \centering
    \begin{tabular*}{\columnwidth}{@{\extracolsep{\fill}}c|ccc} 
    \toprule
    \ &w/o sampling\ &Ours\ \\
    \midrule
    chair(min)\ &169.7\ &\textbf{7.3}\ \\
    car(min)\ &65.2\ &\textbf{7.7}\ \\
    \bottomrule
    \end{tabular*}
    \caption{\textbf{\textit{Running time on whether using sampling.}} Our sampling method can significantly reduce the time cost.}
    \label{tab:time}
\end{table}

%% file: sec/5_conclusion.tex
\section{Conclusion, Limitation and Future Work}
\label{sec5:conclusion}
We introduced CAGE-GS, a novel cage-based deformation framework for 3D Gaussian Splatting (3DGS) that enables high-fidelity shape transformations while preserving fine texture details. By leveraging a learned cage structure from the target shape, our method provides a structured deformation space, ensuring geometric alignment without requiring manual intervention. Additionally, our Jacobian matrix-based optimization effectively updates Gaussian parameters, preventing texture distortion and preserving high-fidelity rendering quality. Extensive experiments demonstrate that CAGE-GS significantly outperforms existing techniques in both efficiency and deformation quality. Its compatibility with various target representations—including texts, images, point clouds, meshes, and 3DGS models—makes it a versatile tool for 3D content creation, scene editing, and shape manipulation.

% Our method is highly compatible, accommodating various model formats as targets, enabling flexible model creation. Experimental results show that our method successfully transfers challenging deformations across multiple datasets. We hope this research inspires further advancements in freeform 3D modeling and precise shape manipulation.

\noindent
\textbf{Limitation.}
Despite its advantages, CAGE-GS has limitations. Our method cannot guarantee that straight lines, plains, and parallel structures remain unchanged during deformation, especially in artificial shapes. Additionally, certain deformations may be better handled using alternative parameterization methods. For example, adjusting only parts of a model might be more efficiently achieved through direct manipulation, such as dragging operations.

\noindent
\textbf{Future work.}
In the future, we aim to design an end-to-end framework for jointly optimizing the Jacobian matrix and Gaussian distribution. Specifically, we plan to integrate neural Jacobian techniques ~\cite{aigerman2022neural} into network, enabling direct learning-based adjustment of Gaussian parameters. Additionally, we intend to extend our method to other applications, such as model registration, animation synthesis, and interactive 3D editing.